\documentclass{jfm}

\input{userDefs/myPreamble}
\input{userDefs/myCommands}

\usepackage{graphicx}
\usepackage{newtxtext}
\usepackage{newtxmath}
\usepackage{natbib}
\usepackage{hyperref}
\hypersetup{
colorlinks = true,
urlcolor   = Blue,
citecolor  = Black,
}

\newcommand{\RomanNumeralCaps}[1]

\title{Taylor rolls on tour:\\Slow drift of turbulent large-scale structures\\in flows with continuous symmetries}
\author{%
Daniel~Feldmann\aff{1}\corresp{\email{daniel.feldmann@zarm.uni-bremen.de}}
\and
Marc~Avila\aff{1}\aff{2}\corresp{\email{marc.avila@zarm.uni-bremen.de}}}
\affiliation{%
\aff{1}University of Bremen, Center of Applied Space Technology and Microgravity (ZARM),\\Am Fallturm 2, 28359 Bremen, Germany.
\aff{2}University of Bremen, MAPEX Center for Materials and Processes,\\Am Biologischen Garten 2, 28359 Bremen, Germany.}

\begin{document}
\maketitle

\begin{abstract}
In \RBC and \TCF cellular patterns emerge at the onset of instability and
persist as large-scale coherent structures in the turbulent regime. Their
long-term dynamics has been thoroughly characterised and modelled for the case
of turbulent convection, whereas turbulent Taylor rolls have received much less
attention. Here we present direct numerical simulations of axisymmetric \TCF and
show a discontinuous phase-transition to spatio-temporal chaos as the system
size increases. Beyond this transition, Taylor rolls suddenly undergo erratic 
drifts evolving on a very slow time scale. We estimate an effective diffusion
coefficient for the drift and compare the dynamics to analogous motions in \RBC
and Poiseuille flow, suggesting that this spontaneous diffusive displacement of
large coherent structures is common among different types of wall-bounded
turbulent flows.
\end{abstract}


\section{Introduction}
\label{sec:introduction}


The smallest eddies in turbulent flows are dictated by the fluid's kinematic
viscosity ($\nu$) and dissipation, whereas the largest ones are shaped by the
flow geometry, boundary conditions and source of driving. Very large coherent
motions in the flow field (superstructures) carry a substantial part of the
kinetic energy, which increases as the Reynolds number (\Reynolds) increases
\citep{Smits2011}.
Understanding their role in transport and mixing is an active field of research,
with many open questions relevant for predicting and modelling environmental
fluid flows \citep{Dauxois2021}. In systems with linear instabilities, such as
\RBC (RBC) and \TCF (TCF), the origin of turbulent superstructures can be traced 
down to the onset of hydrodynamic instability. 
For TCF, Taylor rolls emerge from the primary instability of
circular Couette flow \citep{Taylor1923}, and then undergo a sequence of
bifurcations \citep{Coles1965, Fenstermacher1979, Prigent2006}, which increase
the spatio-temporal complexity of the flow as \Reynolds increases
\citep{Feldmann2023}. Seemingly, they persist in the form of turbulent Taylor
rolls up to the highest \Reynolds investigated to date \citep{Lathrop1992,
Ravelet2010, Huisman2014, Ostilla-Monico2016a, Sacco2019}. We refer to
\citet{Grossmann2016} for a recent review of turbulent \TCF. \par
Turbulent Taylor rolls have been elucidated in experiments and direct numerical
simulations (DNS) employing temporal averages of the velocity field
\citep{Dong2007, Ravelet2010, Ostilla-Monico2013, Huisman2014,
Ostilla-Monico2016, Ostilla-Monico2016a}, based upon the assumptions that the
rolls remain stable and do not travel in axial direction ($z$). In most
laboratory experiments the cylinders are bounded by solid end walls, whereas in
DNS axially periodic boundary conditions (BC) are usually employed. This renders
$z$ homogeneous and enables the usage of short computational domains, which
typically accommodate one or two pairs of Taylor rolls \citep{Dong2007,
Brauckmann2013, Ostilla-Monico2016, Ostilla-Monico2016a, Sacco2019}. In these
computations, typical observation times do not exceed a few hundred convective
time units. This raises the question of whether Taylor rolls remain stable and
stationary up to arbitrarily long times. In cylindrical RBC cells, for example,
the characteristic large-scale circulation (LSC) are known to undergo
spontaneous diffusive meandering in the naturally homogeneous (\ie azimuthal)
direction \citep{Sun2005, Brown2006, Xi2006, Brown2008}. Slow dynamics of the
LSC -- clearly separated from the time scale of the turbulent fluctuations --
was also shown more recently in DNS of rectangular RBC at Rayleigh numbers up to
$\Rayleigh = \num{e7}$ \citep{Pandey2018}. Similarly, \citet{Kreilos2014} found
slow spanwise displacements of velocity streaks in turbulent boundary layer and
Poiseuille flows.\par
In this paper, we reveal a discontinuous phase-transition giving rise to
slow, large-scale dynamics in axisymmetric TCF. Beyond a critical domain size,
spatio-temporal chaos emerges and the Taylor rolls undergo erratic drifts in
$z$. Compared to the cylinder rotation, the drift speeds are small, but large
roll displacements can occur on a slow time scale. We show that the drift
statistics are consistent with a Wiener process and characterise the motion with
an effective diffusion coefficient of the order of $\nu$.

\section{Computer experiments}
\label{sec:computerExperiments}

We perform axisymmetric DNS of \TCF with periodic BC in $z$ for moderate
Reynolds numbers, allowing for both, large computational domains and long
integration times at affordable computing costs. We integrate the incompressible
Navier--Stokes equations (subject to no-slip BC in $r$) forward in time ($t$)
using our pseudo-spectral DNS code \nsc \citep{Lopez2020}. The equations are
formulated in cylindrical coordinates ($r,\theta,z$) and rendered dimensionless
using $d$, $\sfrac{\nu}{d}$ and $\sfrac{d^2}{\nu}$ (\ie the characteristic
viscous time scale of the problem), as unit length, unit speed and unit time,
respectively. In axisymmetric DNS, the $\theta$-dependence is dropped, but all
three velocity components are computed.\par
Motivated by an exact analogy \citep{Eckhardt2020} between two-dimensional RBC
and axisymmetric TCF in the narrow gap limit ($\eta = \sfrac{r_i}{r_o}
\rightarrow 1$), we set $\eta = \num{0.99}$ and vary $\ReS =
\sfrac{Ud}{\nu}$, $\ROmega = \sfrac{2d\Omega}{U}$ and $\Gamma = \sfrac{L_z}{d}$.
Here, $d = r_o-r_i$ is the gap width between the inner ($i$) and outer ($o$)
cylinder, $L_z$ is the axial length, $U = u_{\theta,i}-\eta\,u_{\theta,o}$ and
$\Omega = \sfrac{u_{\theta,o}}{r_o}$, where $u_{\theta,\sfrac{i}{o}}$ denotes
the azimuthal speed of the cylinders. An important response parameter is the
Nusselt number (\NuS), which quantifies the transport of angular momentum across
the fluid layer \citep{Eckhardt2007a, Brauckmann2016, Eckhardt2020}. In a first
set of DNS, we fix all
parameters, but $\Gamma$, to investigate the onset of spatio-temporal chaos with
respect to the lateral domain size. In a second set (compiled in
table~\ref{tab:compareCases}), we fix $\Gamma$ and explore the effect of shear
(\ReS) and rotation (\ROmega). The initial conditions are chosen to trigger the
desired number of Taylor rolls ($N_R$) necessary to maintain their aspect ratio
constant throughout all DNS runs ($\sfrac{N_R}{\Gamma} = 1$). This is important
because the dynamics is known to depend on $\sfrac{N_R}{\Gamma}$
\citep{Ostilla-Monico2015, Ostilla-Monico2016a, Wang2020, Zwirner2020}. The
highest friction Reynolds number ($\ReTau = \sfrac{u_\tau d}{\nu}$, where
$u_\tau$ is the friction velocity at the cylinder walls) measured in all DNS is
\num{408} (table~\ref{tab:compareCases}). The spatial resolution in terms of wall
units (\ie based on \ReTau and denoted by ${}^+$) is at least $\num{0.07} \le
\Delta r^+\le\num{4.03}$ and $\Delta z^+=\num{4.89}$, which is state of the art
in DNS of wall-bounded turbulence \citep{Ostilla-Monico2016,
Ostilla-Monico2016a, Feldmann2021}.

\section{Drift dynamics}
\label{sec:driftDynamics}

\begin{figure}
\centering
\includegraphics[width=1.0\textwidth]{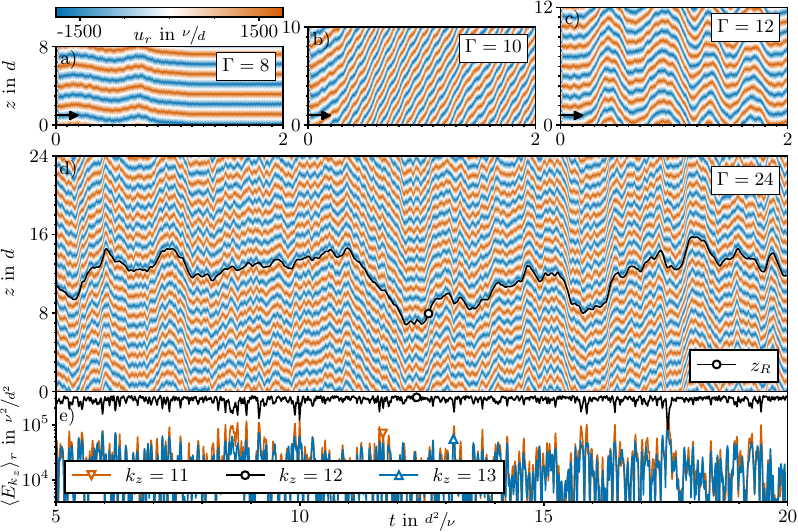}
\caption{Spatio-temporal dynamics of Taylor rolls \wrt the domain size
($\Gamma$). Shown are contours of wall-normal velocities ($u_r$) at mid-gap
extracted from DNS ($\ReS = \num{9475}$, $\ROmega = \num{0.14}$). Arrows
represent \num{2000} convective time units. a): Stationary Taylor rolls. b--d):
Axial motions on a slow time scale. d): Temporal evolution of the phase angle of
the dominant Fourier mode ($k_z = \num{12}$) of $u_r$ on top of the space-time
data as a proxy for the displacement of the rolls ($z_R$) here for twelve pairs
of rolls. e): Kinetic energy time series ($\langle E_{k_{z}}\rangle_r$) for mode
$k_z$.}
\label{fig:spaceTimeEnergyDrift}
\end{figure}

Small domains restrict the dynamics of the system, resulting in nearly
stationary rolls. This is apparent from the space-time diagram of the
wall-normal velocity ($u_r$) for $\Gamma = \num{8}$
(figure~\ref{fig:spaceTimeEnergyDrift}a).
If we now enlarge the domain ($\Gamma \ge \num{10}$, $\sfrac{N_R}{\Gamma} = 1$),
the Taylor rolls undergo large, erratic, collective drifts in $z$, that evolve
on a slow time scale (figure~\ref{fig:spaceTimeEnergyDrift}b--d). In a
domain with $N_R = \num{24}$ rolls, for example, the most energetic axial mode
is always $k_z = \num{12}$ (figure~\ref{fig:spaceTimeEnergyDrift}e), confirming
that the space-time representation of $u_r$ is indeed a robust way to identify
Taylor rolls and to track their dynamics. Every few viscous time units (\eg at
$t\approx\num{17.5}$), the competition with neighbouring modes (here $k_z\in
\{\num{11},\num{13}\}$) represents rare attempts to switch to another state with
eleven or thirteen pairs of rolls. These attempts, however, remained
unsuccessful in all our simulations.\par
\begin{figure}
\centering
\includegraphics[width=0.7\linewidth]{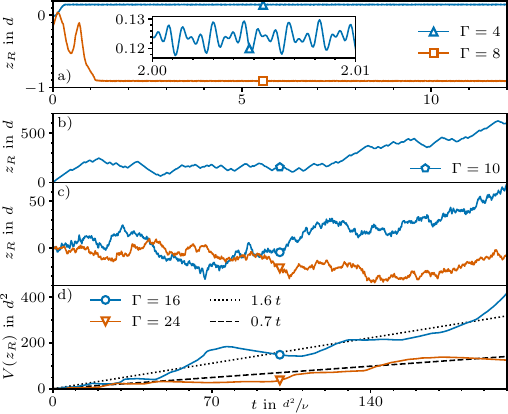}
\caption{Time series of the axial displacement ($z_R$) of Taylor rolls in
different domains ($\Gamma$) extracted from DNS ($\ReS = \num{9475}$, $\ROmega =
\num{0.14}$). a): Chaotic small-scale oscillations. b): Huge drifts. c): Chaotic
oscillations with large erratic drifts. d): Displacement variance ($V(z_R)$) and
linear fit (broken lines).}
\label{fig:driftCompareDomain}
\end{figure}

To analyse the drift dynamics quantitatively, we compute axial Fourier spectra
of $u_r$ (space-time data as in figure~\ref{fig:spaceTimeEnergyDrift}d) and use
the phase angle of the axial dominant mode (here $k_z=12$) to approximate the
displacement of the rolls ($z_R$), as done earlier \citep{Sacco2019}. The
temporal evolution of $z_R$ (figure~\ref{fig:spaceTimeEnergyDrift}d) aligns well
with $u_r$, thereby confirming the suitability of $z_R$ to quantify the drift.
For $\Gamma\le\num{8}$, the rolls first undergo slow transient drifts in the
beginning of the simulation and then ultimately oscillate with tiny amplitudes
and high frequencies about a statistically steady state
(figure~\ref{fig:driftCompareDomain}a).
This fast dynamics of the rolls was
reported earlier for 3d turbulent TCF in a domain accommodating one pair of
rolls \citep{Sacco2019}. By contrast, for $\Gamma = \num{10}$ the rolls tramp
more than $\num{100}d$ before turning back for the first time, and continue
moving erratically thereafter (figure~\ref{fig:driftCompareDomain}b). With
further increasing $\Gamma$, these excursions persist but become less extreme
(figure~\ref{fig:driftCompareDomain}c).

We quantify the Taylor-roll motion statistically by computing the variance of
the axial displacement, $V(z_R) =\langle z_R^{2}\rangle-\langle z_R\rangle^{2}$,
where angled brackets denote temporal averaging. For $\Gamma\le\num{8}$ the fast
dynamics of the rolls is centred around a fixed location and $V(z_R)$ quickly
saturates to a constant, in agreement with \citet{Sacco2019}, who reported
Gaussian fluctuations of $z_R$ with constant variance. By contrast, for
$\Gamma \ge \num{10}$, $V(z_R)$ grows approximately linearly with time, as in a
Wiener process (figure~\ref{fig:driftCompareDomain}d). The drift of the rolls
can be thus characterised with an effective diffusion coefficient ($D_R$),
as the slope of a linear fit to the variance. We generally discard the first
$\num{2}\sfrac{d^2}{\nu}$ to exclude initial transients. 

\section{Discontinuous phase transition}
\label{sec:discontinuousPhaseTransition}

\begin{figure}
\centering
\includegraphics[width=0.7\textwidth]{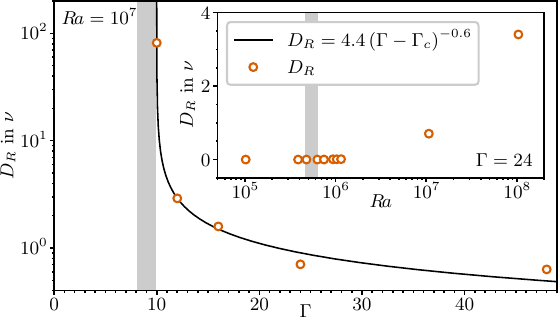}
\caption{Phase transition for the onset of large, erratic drifts \wrt the domain
size ($\Gamma$). Beyond the critical point (grey bar, $\Gamma_c=\num{9.99}$ in
the fit), the motion can be characterised by an effective diffusion coefficient
($D_R$, figure~\ref{fig:driftCompareDomain}d). The inset compares $D_R$ for TCF
at different Rayleigh numbers (\Rayleigh).}
\label{fig:phaseTransitionDiffusion}
\end{figure}

\begin{figure}
\centering
\includegraphics[width=0.7\linewidth]{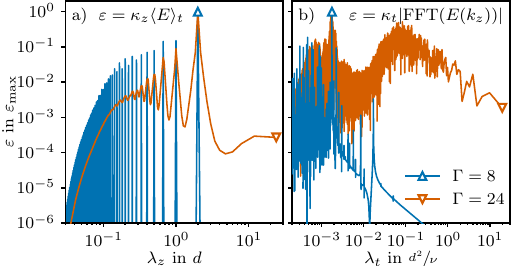}
\caption{Transition from temporal to spatio-temporal chaos in TCF
($\ReS=\num{9475}$, $\ROmega=\num{0.14}$) \wrt the domain length ($\Gamma$). a):
Time averaged, premultiplied axial Fourier spectra of the $u_r$ data
(figure~\ref{fig:spaceTimeEnergyDrift}a,d) versus axial wavelengths $\lambda_z =
\sfrac{2\pi}{\kappa_z}$. b): Premultiplied temporal Fourier spectra of the modal
kinetic energy (figure~\ref{fig:spaceTimeEnergyDrift}e) for the dominant mode
(\eg $k_z=\num{4}$ for $\Gamma=\num{8}$) versus temporal wavelengths $\lambda_t=
\sfrac{2\pi}{\kappa_t}$.}
\label{fig:spectraCompare}
\end{figure}

The dependence of the diffusion coefficient on the domain size is shown in
figure~\ref{fig:phaseTransitionDiffusion}. Our data suggest a divergence of
$D_R$ near a critical point, $\Gamma_c=\num{9.99}$, followed by a monotonous
decrease as $\Gamma$ increases. To examine the nature of this discontinuous
transition, we compare spatial and temporal Fourier spectra from sub- and
super-critical domains (figure~\ref{fig:spectraCompare}). For $\Gamma =\num{8}$,
the axial spectrum of $u_r$ presents discrete peaks at wavelength $\lambda_z=2d$
and its harmonics only (figure~\ref{fig:spectraCompare}a). This implies that the
flow state consists of four perfectly synchronised copies of one pair of Taylor
rolls. In fact, when comparing this state to those obtained for
$\Gamma\in\{2,4\}$, the same Nusselt numbers
($\langle\NuS\rangle=\num{12.82476}$) and spectra (not shown) are recovered. By
contrast, the states obtained for $\Gamma\ge\num{10}$ exhibit continuous spatial
spectra (\eg $\Gamma =\num{24}$ in figure~\ref{fig:spectraCompare}a), indicating
spatial defects in the roll structure; \ie there are no identical rolls in the
entire stack. A similar transition to spatio-temporal chaos was reported before
for axially oscillated \citep{Avila2007} and hydromagnetic \citep{Guseva2015}
TCF. For both, however, no slow large-scale drift of the roll patterns was
reported, possibly due to much shorter simulation times.

The transition to spatio-temporal chaos also alters the temporal spectra
(figure~\ref{fig:spectraCompare}b). For $\Gamma<\Gamma_c$, the temporal spectrum
is continuous, indicating temporal chaos, and exhibits a peak at about \num{20}
convective time units before sharply falling. This peak is associated to the
fast, small-displacement dynamics with $D_R = 0$ reported by \citet{Sacco2019}.
For $\Gamma>\Gamma_c$, the temporal spectrum features an additional broad peak
at about $\num{0.1}\sfrac{d^2}{\nu}$, corresponding to the slow drift dynamics
characterised by a Wiener process. The transition to spatio-temporal chaos is
also reflected in the mean Nusselt number, but only in the third digit ($\langle
\NuS\rangle = \num{12.78}\pm\num{0.04}$ for all $\Gamma$). We note that while
much effort has been dedicated to remove drifts in the analysis of turbulent
dynamics of wall-bounded flows \citep{Willis2013, Budanur2015}, here the onset
of spatio-temporal chaos appears intrinsically linked to the slow, erratic drift
dynamics.

\section{Dependence of the drift dynamics on the flow configuration}
\label{sec:flowConfiguration}

\begin{table}
\centering
\begin{tabular}{ccccccccccc}
\toprule
\Rayleigh & \ReS & \ROmega && \NuS & \Nusselt & \ReTau &&
$V(v_{R})^{\sfrac{1}{2}}$ & $V(\overline{u}_{z})^{\sfrac{1}{2}}$ &
$D_R$ \\
\midrule
\num{e5}&  \num{925}&\num{0.14}&& \num{4.5}&          & \num{32}&&\num{e-14}&\num{e-13}&\num{e-29} \\ 
\num{e6}& \num{2940}&\num{0.14}&& \num{7.4}&          & \num{74}&& \num{0.4}& \num{1.1}&\num{0.008}\\ 
\num{e7}& \num{7138}&\num{0.30}&&\num{10.4}&\num{14.4}&\num{137}&& \num{6.4}& \num{8.8}&\num{0.4}  \\ 
\num{e7}& \num{9475}&\num{0.14}&&\num{12.8}&\num{14.7}&\num{175}&& \num{6.3}& \num{8.8}&\num{0.7}  \\ 
\num{e7}& \num{9475}&\num{0.14}&&\num{12.8}&          &\num{175}&& \num{2.2}&      NOFX&\num{0.07} \\ 
\num{e7}&\num{15779}&\num{0.05}&&\num{14.6}&\num{15.3}&\num{241}&& \num{8.0}&\num{10.5}&\num{2}    \\ 
\num{e8}&\num{29539}&\num{0.14}&&\num{22.4}&          &\num{408}&&\num{12.0}&\num{13.8}&\num{3}    \\ 
\bottomrule
\end{tabular}
\caption{Taylor roll dynamics in different set-ups ($\Gamma=\num{24}$). Listed
are control parameters (\Rayleigh, \ReS, \ROmega), response parameters (\NuS,
\Nusselt, \ReTau), standard deviations of the drift speed ($v_R$), net axial
flux ($\overline{u}_z$), and the effective diffusion coefficient ($D_R$). NOFX
means $\overline{u}_z=0$ enforced.}
\label{tab:compareCases}
\end{table}

We exploit the analogy between TCF and RBC \citep{Bradshaw1969,
Veronis1970, Prigent2006, Eckhardt2007a, Eckhardt2020}, to demonstrate that the 
drift dynamics is found throughout the centrifugally unstable co-rotating
regime.  According to the exact Navier--Stokes mapping of \citet{Eckhardt2020}, 
axisymmetric TCF systems in the narrow-gap limit ($\eta\rightarrow 1$) are
exactly identical if $\Rayleigh=\ReS^2\,\ROmega\left(1-\ROmega\right)=
\text{const.}$; \ie the large-scale drift dynamics is identical as well. Indeed,
for moderate outer cylinder rotation ($\ROmega\in\{\num{0.14},\num{0.30}\}$),
the drift statistics are similar (table~\ref{tab:compareCases}) and the same is
true for the corrected Nusselt number ($\Nusselt = 1+\sfrac{\NuS-1}{1-\ROmega}$)
\citep{Eckhardt2020}. We attribute the small deviations to small, yet finite,
curvature effects ($\eta=\num{0.99}< 1$), which are not included in the analogy.
For very slow outer cylinder rotation ($\ROmega = \num{0.05}$), the drift
statistics deviate noticeably. This is as expected, because the exact analogy
breaks down in the limit of a stationary outer cylinder ($\ROmega = \num{0}$).
Next, we fix \ROmega and vary \ReS. As \ReS increases, the rolls become more
active. Specifically, the variance of the drift speed, $V(v_R)=V(\dot{z}_R)=
\langle\dot{z}_R^{2}\rangle-\langle\dot{z}_R\rangle^{2}$, increases as \ReS
increases (table~\ref{tab:compareCases} and inset in
figure~\ref{fig:phaseTransitionDiffusion}). This is consistent with RBC
experiments \citep{Xi2006}, where the rate of erratic rotation of the LSC 
increases tenfold as \Rayleigh increases from \num{e9} to \num{e10}.

\begin{figure}
\centering
\includegraphics[width=0.7\linewidth]{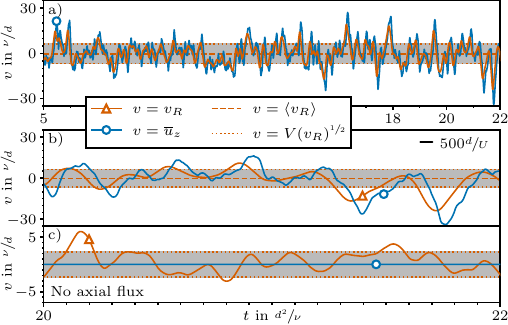}
\caption{Taylor roll dynamics in DNS ($\ReS=\num{9475}$, $\ROmega = \num{0.14}$,
$\Gamma = \num{24}$) with and w/o axial flux constraint. a): Time series of the
drift speed ($v_R$) and the net axial flux ($\overline{u}_z$) for the case in
figure~\ref{fig:spaceTimeEnergyDrift}d. b): Zoom of a). Dash represents
\num{500} convective time units. c): Time series from DNS with $\overline{u}_z =
0$ enforced.}
\label{fig:speedCompareFlux}
\end{figure}

The axial drift of the Taylor rolls is associated to a net mass flux in $z$ with
mean speed $\overline{u}_z$. This flux is strongly correlated to the drift speed
($v_R$) of the rolls (figure~\ref{fig:speedCompareFlux}a), and raises the
question of whether the roll displacement causes the net axial flux or vice
versa. The fact that $v_R$ is approximately $\num{500}\sfrac{d}{U}$ ahead of
$\overline{u}_z$ suggests the former (figure~\ref{fig:speedCompareFlux}b). We
probe this hypothesis by enforcing $\overline{u}_z=\num{0}$, as in laboratory
experiments of TCF with end-walls. In our DNS with axially periodic BC, we
enforce $\overline{u}_z =0$ by imposing an appropriate adverse pressure gradient
at each time-step. This technique was previously applied to successfully compare
axially periodic simulations to lab experiments for \TCF with radial heating
\citep{Ali1990} and axially oscillating inner cylinder \citep{Marques1997}. As a
result of suppressing the axial mass flux, $v_R$, $V(v_R)$ and $D_R$ are
substantially reduced (figure~\ref{fig:speedCompareFlux}c,
table~\ref{tab:compareCases}), but when rescaled, the drift dynamics remains
qualitatively unaltered (figure~\ref{fig:driftCompareSystem}a). Specifically,
$V(z_R)$ still increases linearly with time
(figure~\ref{fig:driftCompareSystem}e), although at a slower pace.

\section{Discussion}
\label{sec:discussion}

\begin{figure}
\centering
\includegraphics[width=0.7\linewidth]{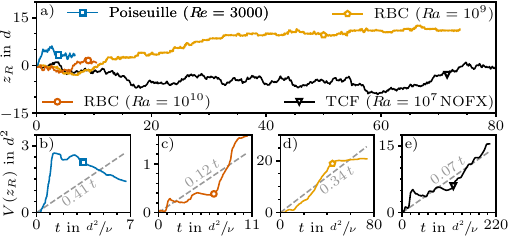}
\caption{Large-scale drift ($z_R$) in different systems. a): Axial Taylor roll
drift in TCF w/o net axial flux (NOFX). Azimuthal meandering of single
convection rolls in cylindrical RBC cells; eleven ($\Rayleigh=\num{e10}$
\citep{Brown2006}) and \num{33} ($\Rayleigh=\num{e9}$ \citep{Xi2006}) days lab
experiments. Spanwise streak displacement in Poiseuille flow DNS
\citep{Kreilos2014}. b--e): Corresponding displacement variance ($V(z_R)$)
including linear fits (broken lines) to estimate an effective diffusion
coefficient.}
\label{fig:driftCompareSystem}
\end{figure}

We have shown that axisymmetric Taylor--Couette flows exhibit a phase transition
to spatio-temporally chaotic Taylor rolls that drift erratically in the axial
direction. At long time scales, the drift motion is diffusive and can lead to
very large displacements. Future works should clarify whether this dynamics
persists in three-dimensional TCF simulations and in experiments with end-walls.
We note that even with walls, flow patterns can drift in $z$ with phase being
created/annihilated near the walls \citep{Heise2008}.\par
The roll displacements extracted from our DNS are compared to the meandering of
the LSC in RBC \citep{Brown2006, Xi2006} and to spanwise streak displacements in
Poiseuille flow \citep{Kreilos2014} in figure~\ref{fig:driftCompareSystem}. For
the sake of comparison, we converted the rotation angle to a length as $z_R(t) =
R\theta(t)$ using the radius ($R$) of the RBC cell. Additionally, we rescaled
all drift signals to the viscous time unit, which is also the relevant one of
the exact analogy \citep{Eckhardt2020}. The qualitative agreement is remarkable
and suggests that this slow dynamics might be inherent to large-scale motions in
many fluid systems. However, longer RBC and Poiseuille flow runs would be needed
to confirm the Wiener statistics found here for TCF, and to estimate the
corresponding diffusion coefficients (figure~\ref{fig:driftCompareSystem}b--e).
Additional statistical analyses and modelling strategies previously applied to
three-dimensional RBC \citep{Brown2007, Brown2008} could help elucidate further
aspects of the drift dynamics reported here and deepen this comparison.\par
We stress that the shortest time series considered here
(figure~\ref{fig:driftCompareDomain}) correspond to $\num{200}\ReS =
\orderof{\num{e6}}$ convective time units and to $\num{200}\sqrt{\Rayleigh} =
\orderof{\num{e5}}$ free-fall time units in RBC ($\Rayleigh = \num{e7}$). These
observation times are comparable to those used to characterise large-scale
states in RBC \citep{Pandey2018, Wang2020}, but they are several orders of
magnitudes longer compared to typical observation times in high-\Reynolds TCF
studies \citep{Brauckmann2013, Huisman2014, Ostilla-Monico2016,
Ostilla-Monico2016a, Sacco2019}. In the Taylor--Couette apparatus of
\citet{Huisman2014}, for example, this would correspond to a measurement time of
two weeks. 



\backsection[Acknowledgements]{We appreciate stimulating discussions with
Alberto Vela-Mart\'in and Daniel Mor\'on and we gratefully acknowledge the
received funding.}

\backsection[Funding]{We received financial support from the German Research
Foundation (DFG) through the priority programme \spp and computational resources
provided by the \hlrn through project \texttt{hbi00041}.}

\backsection[Declaration of interests]{The authors report no conflict of
interest.}

\backsection[Data availability statement]{The data that support the findings of
this study will be made openly available in \pangaea at http://doi.org\dots}

\backsection[Author ORCIDs]{%
Daniel~Feldmann, \url{https://orcid.org/0000-0002-6585-2875};
Marc~Avila, \url{https://orcid.org/0000-0001-5988-1090}}


\bibliographystyle{jfm.bst}
\bibliography{references.bib}

\end{document}